%% file: sigir-25-dl-annotations-frame.tex
\documentclass[sigconf,natbib]{acmart}
\usepackage{sigir-25-dl-annotations}

\input{local-macros}

\begin{document}

\input{sigir-25-dl-annotations-pre}
\input{sigir-25-dl-annotations-part1}
\input{sigir-25-dl-annotations-part2}
\input{sigir-25-dl-annotations-part3}
\input{sigir-25-dl-annotations-sum}

\raggedright
\raggedbottom
\bibliographystyle{ACM-Reference-Format}
\bibliography{sigir-25-dl-annotations-lit}

\end{document}

%% file: local-macros.tex
\definecolor{ggreen}{rgb}{0.061, 0.563, 0.251}

%% file: sigir-25-dl-annotations-pre.tex
\title[Variations in Relevance Judgments and the Shelf Life of Test Collections]{Variations in Relevance Judgments \texorpdfstring{\\}{} and the Shelf Life of Test Collections}

\author[A.\ Parry]{Andrew Parry}
\affiliation{
\institution{University of Glasgow}
\country{United Kingdom}
\city{Glasgow}
}

\author[M.\ Fr{\"o}be]{Maik Fr{\"o}be}
\affiliation{
\institution{Friedrich-Schiller-Universit{\"a}t Jena}
\country{Germany}
\city{Jena}
}

\author[H.\ Scells]{Harrisen Scells}
\affiliation{
\institution{University of T{\"u}bingen}
\country{Germany}
\city{T{\"u}bingen}
}

\author[F.\ Schlatt]{Ferdinand Schlatt}
\affiliation{
\institution{Friedrich-Schiller-Universit{\"a}t Jena}
\country{Germany}
\city{Jena}
}

\author[G.\ Faggioli]{Guglielmo Faggioli}
\affiliation{
\institution{University of Padua}
\country{Italy}
\city{Padua}
}

\author[S.\ Zerhoudi]{Saber Zerhoudi}
\affiliation{
\institution{Universit{\"a}t Passau}
\country{Germany}
\city{Passau}
}

\author[S.\ MacAvaney]{Sean MacAvaney}
\affiliation{
\institution{University of Glasgow}
\country{United Kingdom}
\city{Glasgow}
}

\author[E.\ Yang]{Eugene Yang}
\affiliation{
\institution{Johns Hopkins University}
\country{USA}
\city{Baltimore}
}

\renewcommand{\shortauthors}{Andrew Parry et al.}

\begin{abstract}
The fundamental property of Cranfield-style evaluations, that system rankings are stable even when assessors disagree on individual relevance decisions, was validated on traditional test collections. However, the paradigm shift towards neural retrieval models affected the characteristics of modern test collections, e.g., documents are short, judged with four grades of relevance, and information needs have no descriptions or narratives. Under these changes, it is unclear whether assessor disagreement remains negligible for system comparisons. We investigate this aspect under the additional condition that the few modern test collections are heavily re-used. Given more possible query interpretations due to less formalized information needs, an ``expiration date'' for test collections might be needed if top-effectiveness requires overfitting to a single interpretation of relevance. We run a reproducibility study and re-annotate the relevance judgments of the 2019~TREC Deep Learning track. We can reproduce prior work in the neural retrieval setting, showing that assessor disagreement does not affect system rankings. However, we observe that some models substantially degrade with our new relevance judgments, and some have already reached the effectiveness of humans as rankers, providing evidence that test collections can expire.
\end{abstract}
\keywords{Evaluation; Test collections; Collection Reliability}


\ccsdesc[500]{Information systems~Evaluation of retrieval results}

\setcopyright{acmcopyright}
\copyrightyear{2025}
\acmYear{2025}
\setcopyright{cc}
\setcctype{by}
\acmConference[SIGIR '25]{Proceedings of the 48th International ACM SIGIR Conference on Research and Development in Information Retrieval}{July 13--18, 2025}{Padua, Italy}
\acmBooktitle{Proceedings of the 48th International ACM SIGIR Conference on Research and Development in Information Retrieval (SIGIR '25), July 13--18, 2025, Padua, Italy}\acmDOI{10.1145/3726302.3730308}
\acmISBN{979-8-4007-1592-1/2025/07}

\maketitle

%% file: sigir-25-dl-annotations-part1.tex
\section{Introduction}

TREC-style test collections enable the evaluation and, thereby, the development of retrieval systems. Making these collections robust and reusable is not trivial, partly because of the inherent subjectivity of relevance~\cite{taube:1965}. Assessing the reliability of relevance judgments involves assessing inter-annotator agreement and is comparably expensive~\cite{lesk:1968, burgin:1992, voorhees:2000}. Consequently, only a few studies have investigated how the variability of relevance judgments affects the evaluation of retrieval systems. However, those findings were highly important for the design of retrieval test collections as they showed that even when relevance is subjective and therefore annotators disagree~\cite{schamber:1990, soboroff:2024}, subsequent aggregated effectiveness evaluations~\cite{lesk:1968, burgin:1992} and system rankings~\cite{voorhees:2000} are stable, ensuring the reusability of test collections.

\input{src/table/table-overview-studies}

Contrasting older collections, topics in modern test collections often do not have descriptions or narratives~\cite{craswell:2019}, which is known to affect agreement~\cite{rees:1967, voorhees:1997}, documents are often shorter, and relevance is assessed on a graded scale, which has not been investigated in re-annotation studies. Table~\ref{table-overview-studies} contrasts previous studies and the properties of the corresponding test collections with the TREC Deep Learning 2019~\cite{craswell:2019} (DL'19) test collection we use in our reproducibility work. The changes in modern collections potentially increase assessment variability because an assessor can freely interpret the query intent and context that may be missing from the document to assess its relevance on a fine-grained scale. In this reproducibility study, we investigate if the robustness of classical test collections still holds for modern test collections under these new conditions.

These shifts in test collection design coincide with advances in retrieval methodologies; retrieval pipelines increasingly comprise components where neural language models estimate relevance. These systems have continuously improved over multiple iterations through improved training processes~\cite{pradeep:2022}, through distillation from large language models~\cite{pradeep:2023, schlatt:2024b}, or by directly using large language models for ranking~\cite{sun:2023}. We hypothesize that iterative model improvements on static benchmarks may overfit to the specific opinions expressed in static relevance judgments, introducing systemic bias, which can be characterized as ``bias by research design'' in which the influence of previous works can bias new findings implicitly towards a common benchmark~\cite{hovy:2021}. The TREC DL'19 track remains a commonly used standard for benchmarking systems, being cited and most likely evaluated hundreds of times (Cited 32~times at ECIR'24 and SIGIR'24 alone). \citet{craswell:2021} alluded to this point, warning against the iterative design processes employed in ad-hoc neural ranking.

Thus, we look to replicate the classic setting for evaluating the robustness of collections proposed by \citet{voorhees:2000} for modern collections, with the additional hypothesis that bias in the system orderings may be introduced through model iteration in addition to human disagreement. Since a single expert human assessor's relevance grades on a given topic are subjective and ``equally plausible'' as another assessor's, we can compare the effectiveness of systems using multiple ``equally correct'' sets of relevance judgments. If a system's relative effectiveness degrades when evaluated using alternative judgments, it may indicate that models overfit the preferences expressed by the collection's official judgments.

In this paper, we conduct a re-annotation of TREC DL'19 using two ``secondary'' annotators for every topic. Based on our new relevance judgments, we first provide a qualitative analysis of this process and validate the findings of previous work in a modern setting. That is, we observe low agreement among assessors but system rankings remain stable. Secondly, treating human annotators as manual TREC runs, we find that one can no longer meaningfully discriminate between the state-of-the-art and human judges on this collection. A reasonable ``upper bound'' effectiveness is around nDCG@10 of 0.81 (due to ambiguities in relevance assessment). We suggest that reported results that are higher than this bound on DL'19 may already overfit to the relevance labels. Note that we are not aware of any existing system reporting results exceeding this bound.%
\footnote{We do not consider large ensemble systems, given their risks~\cite{NetflixPrize}.} 
However, the leading systems today are very close, with RankZephyr~\cite{pradeep:2023} and RankGPT~\cite{sun:2023} reaching between 0.78--0.80~nDCG@10 with a sufficiently strong first stage. We additionally find that closed-sourced large language models and systems distilled from them can fluctuate greatly in relative effectiveness, often degrading system rank on our new relevance judgments.

Our reproducibility effort spawned from the ECIR~2024 Collab-a-thon on Neural IR~\cite{macavaney:2024}. A subset of the IR community discussed that the current generation of retrieval models may have reached a practical upper bound on existing test collections, but we were unsure when to consider a test collection to be saturated or expired. Our study reproduced prior findings on the inherent subjectivity and variability of relevance judgments (especially in the absence of a well-defined narrative), and it established an empirical upper bound on TREC DL'19. Therefore, we encourage caution when attempting to maximize a given metric as we approach and exceed human bounds on relevance estimation. Exceeding human effectiveness, given the subjectivity of relevance, may not lead to an optimal system but instead stagnation of progress in the field. We publish our code and relevance judgments for reproducibility and to encourage future studies.\footnote{\url{https://github.com/Parry-Parry/sigir25-annotation}}

%% file: src/table/table-overview-studies.tex
\begin{table}[tb]
    \centering
    \setlength{\tabcolsep}{2pt}
    \footnotesize
    \caption{Overview of annotator agreement studies that we reproduce in the TREC Deep Learning scenario. We contrast information needs (length of title, description, narrative), documents (domain and length), evaluations (measure and relevance grades) and observed results of each study (we show the agreement as overlap and the system correlations as $\tau$).}
    \begin{tabular}{@{}l@{}ccc@{\hspace*{3\tabcolsep}}cc@{\hspace*{3\tabcolsep}}cc@{\hspace*{3\tabcolsep}}cc@{}}
        \toprule
        \bfseries Study & \multicolumn{3}{c}{\bfseries Information Need \phantom{0}} & \multicolumn{2}{c}{\bfseries Docs. \phantom{0}} & \multicolumn{2}{c}{\bfseries Evaluation \phantom{0}} & \multicolumn{2}{c}{\bfseries Results \phantom{0}}  \\
        \cmidrule(r{2\tabcolsep}){2-4}
        \cmidrule(r{2\tabcolsep}){5-6}
        \cmidrule(r{2\tabcolsep}){7-8}
        \cmidrule{9-10}
         & Title & Desc. & Narr.  & Dom. & Len. & Meas. & Rel. & Agr. & Corr. \\
        \midrule
        Lesk, Salton~\cite{lesk:1968} & $\times$ & 100.0 & $\times$ & Medical & 103.7 & Prec. & 0--1 & $0.30$ & $\times$ \\
        Burgin~\cite{burgin:1992} & $\times$ & \phantom{0}30.4 & $\times$ & IR & \phantom{0}15.7 & MAP & 0--1 & $0.49$ & $\times$ \\
        Voorhees~\cite{voorhees:2000} & 2.6 & \phantom{0}20.4 & 64.7 & News & 550.5 & MAP & 0--1 & $0.36$ & $0.890$\\
        \midrule
        Our Reproduction & 8.3 & $\times$ & $\times$ & QA & \phantom{0}30.4 & nDCG & 0--3 & $0.15$ & $0.897$\\
        \bottomrule
    \end{tabular}
    \label{table-overview-studies}
\end{table}

%% file: sigir-25-dl-annotations-part2.tex
\section{Related Work}

This section outlines evaluation frameworks in retrieval, previous efforts to assess collection reliability, and judgment stability.

\paragraph{Offline Evaluation}
The Cranfield experiments introduced the modern offline test collection in ad-hoc ranking; experts curated a set of queries, and texts were retrieved from a static collection and exhaustively annotated, allowing for offline relevance judgments~\cite{cleverdon:1966}. Completeness of relevance judgments, i.e., the exhaustive annotation of corpora, is a key factor in the reliability of an offline test collection~\cite{voorhees:2019}; however, the scale of modern corpora makes such a task infeasible. As such, TREC-style collections are collated by pooling the runs of multiple diverse systems to create a set of query--document pairs to be annotated under the assumption that unpooled documents are not relevant~\cite{voorhees:1998}. Assessors then label these pairs and assign a grade of relevance. A single annotator is often assigned to each topic; this decision generally reduces costs and has been validated by several studies measuring agreement between annotators as an effect on the reliability of a collection.

Relevance judgments are subjective and, therefore, can vary among annotators~\cite{mizzaro:1997}, with low inter-annotator agreement~\cite{sakai:2019}, which can indicate low collection reliability. As such, narratives and training are usually provided to annotators describing the information need underlying a query~\cite{harman:2012}. We consider that these homogenized judgments may be a weak point for ongoing collection reliability as we learn from them in a semi-supervised fashion through effective systems acting as teachers.

\paragraph{Notions of Relevance}
The reliability of offline test collections is contested due to the intrinsic ambiguity and biases present in relevance assessment under different human annotators~\cite{taube:1965, regazzi:1988}, annotation settings~\cite{rees:1967}, relevance grading~\cite{cuadra:1967}, and, more recently, the process of collecting candidate documents via pooling~\cite{zobel:1998}. As these concerns have been raised, one method to validate offline test collections is to re-annotate judgments. \citet{lesk:1968} and \citet{burgin:1992} empirically validated the collection reliability under annotator disagreement, observing that system order is maintained under different annotators by measuring variance in precision and recall. When web-scale corpora rendered exhaustive annotation impossible, depth pooling became popularized to provide a minimal set of documents that should be annotated to discriminate system effectiveness. Concerns were two-fold: that systems that contributed to the pool would have an advantage over new systems and that the reduced annotations may disrupt the stability of system orderings. \citet{voorhees:2000} found that system order stability could be maintained under the re-annotation of test collection pools from TREC-4~\cite{harman:1995} and -6~\cite{voorhees:1997}.
\label{sec:factors}%

\paragraph{Collection Retirement}
Previous work has discussed the retirement of a test collection in recent years, though not retired before dissemination the use of TREC 2021 Deep Learning test collection~\cite{craswell:2021} is discouraged by NIST due to the large increase in corpora size leading to pooling bias; both NIST~\cite{voorhees:2022} and participants in the track~\cite{kamps:2021} had concerns regarding the ability to measure recall due to incomplete labeling. However, we are not aware of any work that considers the effect of collection popularity and subsequent bias induced by fitting to relevance judgments over time; our focus on precision is in line with the philosophy of neural systems and the Deep Learning track.

\citet{craswell:2021} warns against using TREC Deep Learning test collections for experimental iteration to reduce the chance of overfitting to judgments. Though this guidance is generally followed with the MSMARCO-dev set acting as a suitable alternative, the broad use of existing strong systems validated on TREC Deep Learning 2019~\citep{hofstatter:2020, wang:2023, pradeep:2023} may lead to overfitting via semi-supervision. We draw a parallel with the work of \citet{armstrong:2009}, which warned that progress might stagnate due to poor practice in evaluation. Similarly, we note that the continued comparison to a dataset that guides our architecture and training pipeline may lead to stagnation in system development.

Within a modern ranker training pipeline, several components depend on existing models for effectiveness. Hard negative mining is a process in which, instead of random negative examples in a contrastive loss, samples are taken from the top most relevant documents to a query with the assumption that the majority will be non-relevant~\cite{karpukhin:2020}. It is common to employ one or several models' pooled rankings to sample negatives~\cite{shen:2022, wu:2023}, with these models often chosen based on their in-domain effectiveness. More explicitly, pipelines are frequently composed of multiple stages of distillation~\cite{wang:2022, wu:2023, schlatt:2024}, which is a semi-supervised process in which a student model is optimized to approximate the output of a teacher model~\cite{hinton:2015}. Again, these models will be chosen based on in-domain effectiveness and coupled with hard negatives. 

%% file: sigir-25-dl-annotations-part3.tex
\section{Evaluating Reliability through Re-Annotation}

The reliability of a test collection, given the subjectivity of relevance, is often validated by re-annotation~\cite{lesk:1968, burgin:1992, voorhees:2000}. Modern IR models are often optimized to mimic previously effective systems through data-driven processes. We examine how these models, designed after a collection's release, perform on new judgments following the process of \citet{voorhees:2000}.

\subsection{Reproduction Source}
We describe the original annotation process applied by \citet{voorhees:2000}. This work was carried out as a component of the curation of the TREC-4~\citep{harman:1995} and TREC-6~\citep{voorhees:1997} corpora. As is standard in the curation of pooled TREC corpora, one ``primary" annotator was assigned to each topic, subsequently judging each document in the pool. Two additional annotators were assigned to each topic to validate the effect of re-annotation on depth-pooled corpora. A sub-sample of judgments was taken from the pool and re-annotated. In TREC-4, a maximum of 200 relevant documents per topic were taken as a secondary pool for re-annotation. Additionally, 200 non-relevant documents were sampled per topic. A binary judgment was assigned to each query--document pair, relevant or non-relevant. The texts constituting the primary annotation pool, left unjudged in the secondary pool, are included in all system evaluations, maintaining identical pooling biases. The comparison of ``secondary" annotations was carried out to assess hypotheses that resurface frequently in IR literature.  This work focuses on the subjectivity of relevance and how pooling affects this subjectivity. Thus, experiments were conducted over a depth-pooled corpus under construction at the time. This process produces a subset of judgments where each query--document pair was re-annotated twice.
\label{sec:perm}

Where possible, we follow the evaluation process of \citet{voorhees:2000}, including analysis of the probability of two systems swapping in effectiveness under different annotators. We define, in an annotation set $A$, an annotation $a\in A$ as a tuple $\{q, d, r\}$ composed of query $q$, document $d$, and judgment $r \in [0, 1, 2, 3]$. For a given annotator pair and a primary annotator, assigned $n$ query-document pairs, we have three annotation sets $A_1, A_2, A_3$ where $A_i=\{a_j\}_{j=0}^n$. For $m$ queries, as noted by \citet{voorhees:2000}, there are $3^m$ possible combinations. Thus, these combinations allow for the greater exploration of judgments that could have been produced from a given annotation process. From the 3 annotation sets, a combination $A^*$ is sampled $s$ times, and systems are evaluated against each combination. The probability of a swap between systems $i$ and $j$ can be evaluated under a given measure $f$ given judgments $A^*$. For systems $S$, we produce a matrix $B\in \mathbb{Z}^{|S|\times|S|}$ where $B[i, j]$ is the number of annotation samples which caused systems $S_i$ to have higher effectiveness than $S_j$. The swap probability between these systems can then be computed as $\frac{\text{min}(B[i, j], B[j, i])}{s}$. Due to the symmetric treatment of system comparisons, this computation is bounded between $0$ and $0.5$.

\subsection{Re-Annotation of a Modern Collection}

While the original study focused on binary judgments, our work adapts this methodology to graded relevance and neural systems under a modern evaluation setting. Our reproduction is conducted on the TREC Deep Learning 2019 track (DL'19)~\cite{craswell:2019}. Specifically, we use the passage collection, which is generally more popular. The proposal of more granular measures that explicitly consider the ordering of a ranking, such as discounted cumulative gain, as opposed to set-based measures, such as Precision@$k$, generally requires graded relevance as opposed to the binary relevance grades of older collections. DL'19 was annotated under this setting with four grades of relevance: `perfectly relevant,' `highly relevant,' `related', and `non-relevant.' This distinction leads to some key differences in our process and analysis.

We treat the original judgments of DL'19 as those of the ``primary'' annotator. In our study, two annotators from a pool of 8 retrieval academics\footnote{These academics contributed to subsequent analysis and write-up in this work. This decision has minimal effect on our findings; nevertheless, we acknowledge this overlap.} are assigned to each topic and act as ``secondary'' annotators (corresponding to 4 topic groups), aligning with the original study in that each topic is annotated by two new annotators; however, the original study does not specify the number of total annotators. The size of the annotator pool used in this work is similar to that of TREC collections~\cite{soboroff:2024}, albeit with a reduced total annotation budget due to cost and following in the sampling setting of \citet{voorhees:2000}. We create a secondary pool composed of all documents in the pool with a grade of 1 or more, and sample 100 documents per annotator from the non-relevant documents of the primary pool. We reduce the number of non-relevant judgments from the pool, motivated by the hypothesis that agreement is far greater over what is non-relevant; for set-based measures, this choice leads to a balance between relevant and non-relevant texts and has minimal effect on nDCG.

Our process for pool creation uses all judged documents at least related to a topic and assigns topics to annotators in a round-robin fashion to balance annotation load while maintaining two annotators per topic. A priority queue of annotator pairs ordered by their current annotation load is produced and assigned the next topic with the greatest number of judgments. The load on each annotator is then balanced by swapping large topics to annotators with minimal load in a second round-robin. We then assign non-relevant texts to each annotator pair with a total mean annotation load of 1125 pairs per annotator.

\paragraph{Annotator Instruction.}
Each annotator was provided with annotation guidelines, including the original relevance grade descriptions from DL'19. Annotators were made aware that instances had been previously annotated for DL'19; however, they were not provided with relevance grades or a grade distribution. Due to the specialist knowledge required to assess relevance in, for example, medical topics, annotators were allowed to familiarise themselves with topics at their discretion. To facilitate reproducibility and prevent exposure to discussion and instances of DL'19 relevance judgments, which may be indexed on common search engines, we instead enforce the use of the ChatNoir search endpoint~\cite{potthast:2012}, which indexes ClueWeb22~\cite{overwijk:2022} and retrieves using BM25.

\subsection{Further Investigation}

Beyond our reproduction setting, we utilise re-annotated query-document pairs to investigate other aspects of collection reliability.

\paragraph{judgment Aggregation}
\citet{voorhees:2000} apply several aggregation operations including combination as described in Section \ref{sec:perm} and aggregate operators. We perform maximum, minimum, and mean pooling over judgments. In the case of mean pooling, we do not perform further quantization; for example, three texts of grades 1, 1.5, and 2, where the second text was annotated 1 and 2 by different annotators, would be compared in that order.

\paragraph{Absent Narratives.}
The absence of narratives or larger descriptions for the TREC Deep Learning collections allows an additional focus for our investigation: Annotators were prompted to note cases of ambiguity and, more generally, their difficulties in annotation. Furthermore, annotators were asked to create their narratives for each query so that agreement among annotators could be investigated with more granularity in the presence of a natural language description of their thought process during annotation. In doing so, we measure query intent as a factor in agreement.

To ablate this factor, we fix narratives after the curation of judgments to assess how narrative interpretation can affect agreement and downstream measures of effectiveness. We take three topics of low, median, and high agreement measured via Fleiss $\kappa$ between the primary and secondary annotators; half of the annotators reference one narrative, and half reference the other. We can then further isolate sources of disagreement to query intent and intrinsic ambiguity of a query title, both qualitatively and quantitatively.

\paragraph{Annotators as Oracles.}\label{sec:oracle}
User models underlie many common IR measures such as discounted cumulative gain~\cite{jarvelin:2002} and mean reciprocal rank (@$k$)~\cite{chappelle:2009}. In performing offline evaluations of search systems using these measures, we aim to model effectiveness via user behavior approximated by relevance judgments. In principle, an optimal system would approach a score of 1 for a normalized metric (assuming $k$ relevant documents exist). However, given the subjective and variable nature of relevance under different query intents, such effectiveness, though possible, may not be indicative of a truly effective system. Human annotators outside the original annotation setting may represent strong search systems as they have exhaustively annotated available documents to some cutoff~$k$. As such, we measure the performance of each annotator under original relevance judgments as an indicator of how non-pooled modern systems are performing relative to a more realistic oracle. Similarly to idealized discounted cumulative gain, we sort by relevance grade post-aggregation. Though graded relevance lacks granularity in contrast to scalar relevance scores, through in-grade permutations (shuffling), we validate that variations in downstream measures of effectiveness are minimal (SD < 0.0001).

\section{Evaluation}

In this section, we outline our investigation and analyze findings when evaluating using secondary annotations.

\subsection{Experiment Design}

Following the re-annotation process of \citet{voorhees:2000}, we observe the impact of re-annotation on the DL'19 test collection, using combinations to assess modern system bias.

\paragraph{Dataset.}
The MSMARCO passage corpus is a collection of around 8.8 million segmented documents covering a broad range of ad-hoc search topics~\cite{nguyen:2016}; this collection and accompanying training topics were mined and annotated from Bing query logs. We re-annotate the TREC Deep Learning 2019 test collection~\cite{craswell:2019}, composed of 43 topics. Unlike the development split released originally with MSMARCO, the DL'19 test set uses densely annotated depth-pooled topics taken from the associated TREC track that year.

\paragraph{Pool Description.}
The DL'19 track primarily focused on the comparison of neural systems in ad-hoc ranking. For equivalence to the re-ranking task, the top 100 texts of all submitted systems from both re-ranking and retrieval tasks were included in our annotation pool. The top 10 texts from each of the 75 systems (44\% Neural LM, 27\% Neural, 29\% lexical/traditional) were added to our pool, and HiCal~\citep{abualsaud:2018} is used to collect an additional 100 texts per topic. Our approach does not annotate all texts and thus uses a subset of this pool.

\input{src/table/models}

\paragraph{Models.}
We compare several neural architectures and training settings. In addition to models that contributed to the original annotation pool, we include systems that use other models for either data augmentation or semi-supervised training via distillation. Table~\ref{tab:models} illustrates the degree of teacher distillation used in each architecture; ColBERT and SPLADE act as end-to-end retrieval or first-stage retrievers, whereas all others re-rank BM25~\cite{robertson:1995} or ColBERT as noted. For monoT5, monoELECTRA, and SetEncoder, we include multiple model sizes, base and large for BERT/ELECTRA variants and base and 3B for T5. For monoELECTRA, we do not use the original checkpoints released by~\citet{pradeep:2022}; we instead use distilled checkpoints released by~\citet{schlatt:2024b} as we are interested in systems that use modern processes. When using RankGPT, we employ GPT-4, GPT-4 Turbo, and GPT-4o.

\paragraph{Measures.}
\citet{voorhees:2000} measure inter-annotator agreement in a pair-wise setting using the overlap of judgments defined as the number of common judgments divided by the total number of judgments. We additionally use Cohen's $\kappa$ over both 4-grade and binarized relevance and Fleiss' $\kappa$, allowing for comparisons between 3 annotators (1 primary, 2 secondary) simultaneously. In the TREC-4 and 6 test collections, mean-averaged precision (MAP) was the primary measure of effectiveness, and therefore, the previous study measured systems by MAP. The TREC DL tracks with four grades of relevance instead primarily measure nDCG@10, and thus, our study is conducted over this metric. Aligned with \citet{voorhees:2000}, we measure downstream system order correlation between judgment sets using measure Kendall's $\tau$; additionally, we measure Spearman's $\rho$ and RBO.

\paragraph{System Order Robustness.}
As we have multiple annotators per query, we can observe pooled judgments across several hypothetical sets; we take all possible combinations of annotators and evaluate both pooled and new models to observe the stability of system order under new judgments. For a given judgment set, we produce a ranking of systems by a target metric (i.e., nDCG@10) and compare this system order to the system ordering under the original judgments. We aggregate rank $\Delta$ for all possible natural combinations over each system to observe how frequently a system degrades over other assessed systems. We measure significant differences in retrieval effectiveness via a paired t-test with Bonferroni correction for different annotator combinations. We measure significant changes in system rank by a Wilcoxon signed test comparing annotator combinations with the system order by official TREC judgments.

\subsection{Pilot Study Findings}
\enlargethispage{2\baselineskip}
We conducted a pilot study over the top 10 pooled texts for 10 randomly sampled topics. Table~\ref{tab:pilot} summarizes the results. We observe that the prevalence of grades 1 and 3 is inflated in these annotations compared to the original annotations. In discussions with annotators, there was ambiguity about what separates a relevant text from a highly relevant text. Additionally, cultural context leads to ambiguity in interpreting a query; for example, a query ``dog day afternoon meaning'' was difficult to interpret without broader context, as a film exists with that name. The query could be interpreted as the meaning of the phrase or the film. From these ambiguities, we allowed annotators to use a controlled search engine to familiarise themselves with the topic. Annotators not from the USA generally noted the difficulty in USA-centric topics; however, we leave bias and query difficulty by nationality to future work, as this line of study is not central to our hypotheses.

\subsection{Core Assessment of Agreement}

\input{src/table/pilot}

\input{src/table/agreement}

Quantification of relevance is confounded by many factors with several works reporting what is generally considered low agreement among annotators~\citep{lesk:1968, burgin:1992, voorhees:2000}. \citet{voorhees:2000} report overlap among annotators as a measure of agreement, observing values around~$0.4$. In Table~\ref{tab:agree}, we present agreement measurements over both 4-grade and binarized relevance. We observe that under 4-grade relevance, all agreement values degrade relative to those stated in previous studies. Fleiss' $\kappa$ measured over primary and secondary annotators indicates near-random agreement but generally improves by around 10 points in all cases over binary judgments. Overlap values over binary judgments generally correspond to those observed by \citet{voorhees:2000}, and over solely secondary annotators, we observe high overlap values, suggesting that while disagreement stems from the subjectivity of relevance and the annotation setting, the process of annotating over 4 grades is generally more variable regardless of topic granularity.  In the third group, we find the lowest values in both 4 grade and binary relevance, which is again reflected in low overlap and Cohen's $\kappa$ values, potentially suggesting diverging query intents or that this allocation of queries was particularly ambiguous. However, we see increases in overlap values when measured solely over secondary annotators, again suggesting that the annotation setting under which judgments are collected contributes to the improved agreement, as noted by \citet{cuadra:1967} and \citet{regazzi:1988} in previous annotation studies. Values of $\kappa$ are low in all cases, suggesting that generally quantifying relevance over 4 grades is more challenging for annotators than the re-annotation setting of \citet{voorhees:2000}. The lack of pre-defined narratives clarifying intent means annotators could interpret a query as they please, potentially resulting in three diverging topic interpretations.

\begin{figure}[t]
\centering
    \includegraphics[width=\columnwidth]{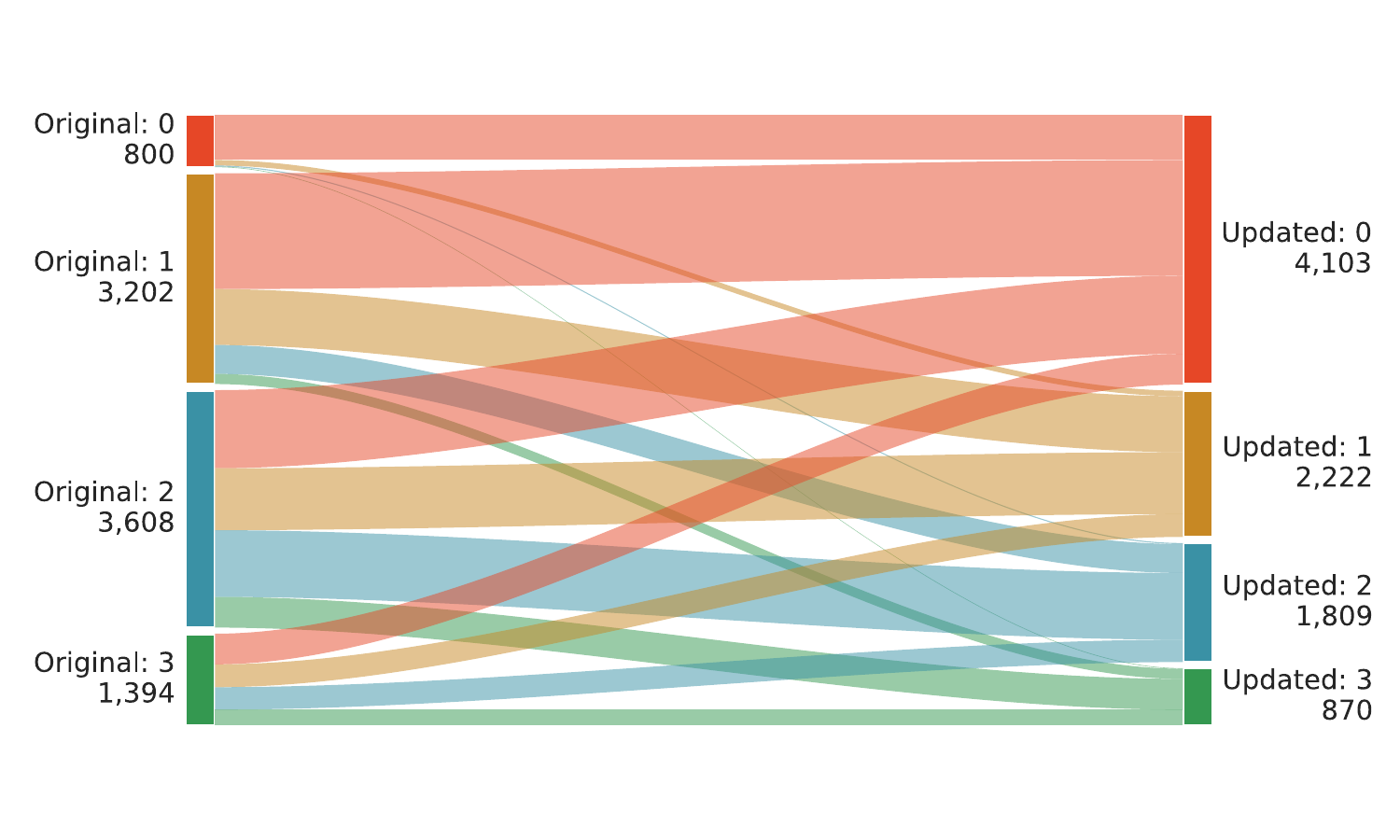}
\caption{Transitions in relevance grades from primary to secondary annotations.}
\Description{A figure showing the flow of grades from primary to secondary annotators, the key observations are that judgments generally reduce e.g. 1 to 0, 2 to 1..}
\vspace{-1em}
\label{fig:grades}
\end{figure}

In Figure \ref{fig:grades}, observe a Sankey flow diagram between the original relevance grade assigned by primary annotators and the new grades by secondary annotators. Like previous work, a secondary annotation setting often leads to the reduction of relevance grades, e.g., 1 to 0 or 2 to 1. Validating our decision to sample a subset of non-relevant judgments, 13\% of 800 non-relevant texts change grade, with 77\% being a change to a grade of 1; we analyse outlier cases (providing a grade of 2 or more) in Section \ref{sec:spec}. Measuring frequencies agnostic of previous judgments, grades 1 and 2 are annotated similarly across annotators, while grades 0 and 3 were annotated with greater variance. This variance in the case of grade 0 is generally consistent across annotators, and we provide a deeper analysis of grade 3 due to varying notions of relevance below.

\subsection{Narratives, Intent, and Relevance}

The observations of grade transitions in our work and generally low agreement noted by prior work, even in a binary setting, may be attributed to several factors noted in Section \ref{sec:factors}. Though we cannot fully isolate factors in the subjectivity of relevance, we choose to investigate the effect of narratives on inter-annotator agreement, given that recent corpora often do not have fixed narratives~\cite{craswell:2019, craswell:2020}, unlike those used in prior re-annotation studies~\cite{harman:1995}. We select topics by median, minimum, and max Fleiss' $\kappa$ values from the annotation pool and consider where disagreement occurs and how narratives differ. Chosen examples illustrate different aspects of agreement, differing query intents, and a narrative's specificity.

The first, taken from Group 1, is a technical query in the medical domain with a Fleiss $\kappa$ of 0.06, indicating disagreement near random guessing. The query ``types of dysarthria from cerebral palsy'' was described by one annotator as a query from a sibling of someone with cerebral palsy looking to understand how their sibling would be affected by their condition in the future. The other annotator described a medical context in which the definition of all conditions is known, and therefore, solely the connection between dysarthria and cerebral palsy was required. The second annotator stated a requirement that relevance was determined by the types and specificity of provided types of dysarthria connected to cerebral palsy, with all other documents being non-relevant. Though both are valid information needs, they represent vastly different query intents, with one annotator's grade frequency matching the overall annotation pool and the other stricter narrative leading to near-total non-relevance.

The second, taken from Group 4, is a more general query, ``is cdg airport in main paris'' with a Fleiss $\kappa$ of 0.08, again indicating low agreement. However, both annotators described a need for exact distances from the airport to either the center of Paris or, more generally, the city of Paris. Both annotators required a succinct description of the distance or travel times to the airport from Paris to be considered perfectly relevant. To be highly relevant, both considered any mention of either travel time or distance measurements. Intents then diverged for partially relevant documents, with one annotator allowing for the mention of costs and other aspects of travel to and from the airport.
In contrast, the other annotator required the airport's position, at minimum, mentioning that it is situated with respect to Paris. Non-relevance was denoted by any other mention of the airport or mentions of distances to destinations near Paris. We find larger differences in lower relevance grades, with one annotator considering the majority of pooled documents to be non-relevant when there is no mention of travel details. In contrast, the other considers them to be relevant because they mention that the airport is situated near Paris in a generic manner.
\label{sec:spec}
The specificity of a query generally leads to lower agreement in our observations. While queries in domains such as law or medicine have a lower overall agreement, we still find cases where generic queries have low agreement due to the interpretations of information needs.

We analyze outlier cases of texts initially labeled non-relevant that, when evaluated under an alternative query intent, were deemed perfectly relevant. In the 9 cases of this, there are cases where a reference is made to the entity of focus, for example ``goldfish'' in the query, ``do goldfish grow'' but only implicitly, for example ``it'' or ``they''. The query was satisfied by a given document if the context is inferred by the surrounding spans of text or from knowing that the document has been retrieved by some system (this is inevitable under pooling), but the information is not self-contained. In some cases, annotators explicitly stated that they considered a document relevant, but when the information is not self-contained, it cannot be annotated as relevant; others took this approach without note.

Overall, many annotation disagreements can occur not only in the high-level query intent of the narrative but also in descriptions of relevance. A diagnostic tool and query-specific grade specifications could be helpful in future work.

\begin{figure*}
    \centering
    \includegraphics[width=\linewidth]{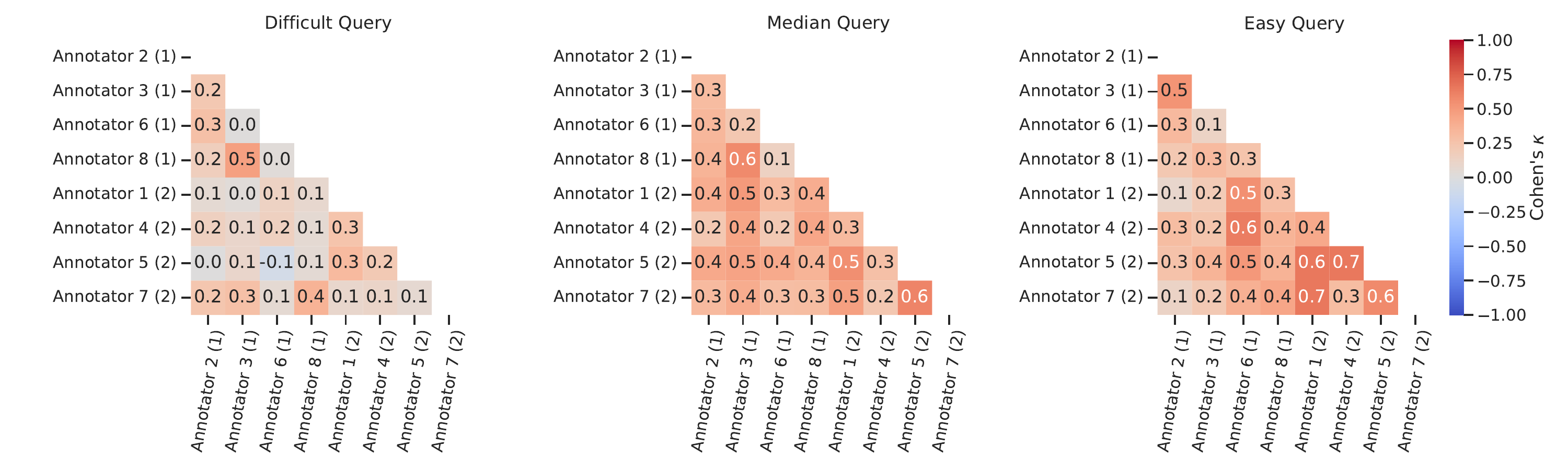}
    \caption{Comparing 3 queries of varying difficulty by inter-annotator agreement under fixed narratives (2 narratives (1) and (2)). Annotators were divided into two groups and were assigned identical document pools. Color indicates strength of agreement; redder indicates stronger agreement, bluer indicates weaker agreement.}
    \Description{A Figure showing agreement, it can be seen that under a more ambiguous query, even with a fixed narrative, agreement can be near random.}
    \label{fig:kappa-diff}
\end{figure*}

\subsection{Narrative and Intent Ambiguity}

Given the lack of public narratives provided in the release of TREC Deep Learning 2019, it is infeasible to isolate query intent as a factor in reliability without first collecting narratives. In fixing narratives, we produce an additional set of relevance judgments for queries of low, median, and high agreement by Fleiss $\kappa$, including the primary annotator. In Figure \ref{fig:kappa-diff}, We classify these queries as difficult, median and easy with respect to query intent determination and measure Cohen's $\kappa$ between the judgments of both narrative groupings. It is clear that a fixed narrative reduces ambiguity with Cohen's $\kappa$, improving by 24\% and 46\% for each annotation group between the difficult and median query; however, between the median and easy query, a decrease of 8\% and an increase of 19\% is observed. 

In several cases, high agreement across different narrative groups occurs; in each case, narratives are of similar detail, defining what intent the user had in searching as well as what should be labelled relevant; we consider that one factor in relevance annotation beyond random noise may therefore be interpretation of a narrative which in the case of a primary annotator is of minimal concern but in human-in-the-loop annotation with generative systems may lead to implications for control as ultimately we can only express our need or a demonstration to the system in natural language. When computing global improvements in agreement, we measure a 79\% increase in agreement between the difficult and median query and an additional 1\% increase comparing the median and easy query. From these observations, we propose that the greatest factor in disagreement is simply the ambiguity of the natural language used to express a query regardless of the narrative applied. Whether or not this is only inherent in pooled corpora would require further investigation as in a setting such as that of TREC DL'19, the ad-hoc realisation of a query intent will be biased by the top-$k$ texts chosen by pooled systems.

\subsection{Core Assessment of System Order}
\begin{figure}
    \centering
    \includegraphics[width=\linewidth]{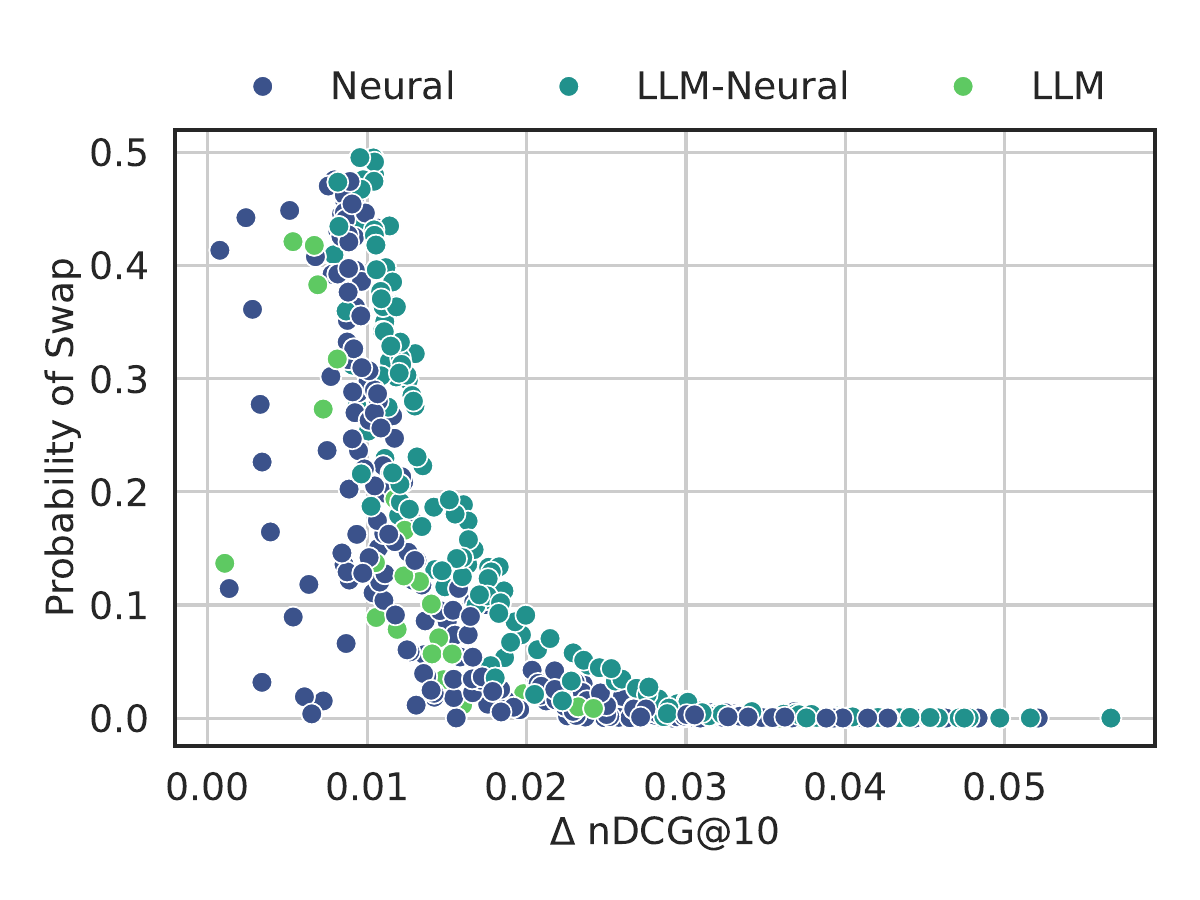}
    \caption{Comparing the stability of judgments through nDCG@10 and system swap probabilities over 10000 judgment combinations. We filter lexical comparisons to improve visibility in the central mass. `LLM-Neural' denotes that a pair contains one LLM-based and one Neural-based ranker.}
    \Description{Shows that LLM and Neural comparisons are frequently unstable under combinations.}
    \label{fig:sample_delta}
\end{figure}
\begin{figure}[t]
\centering
\includegraphics[width=1.1\columnwidth]{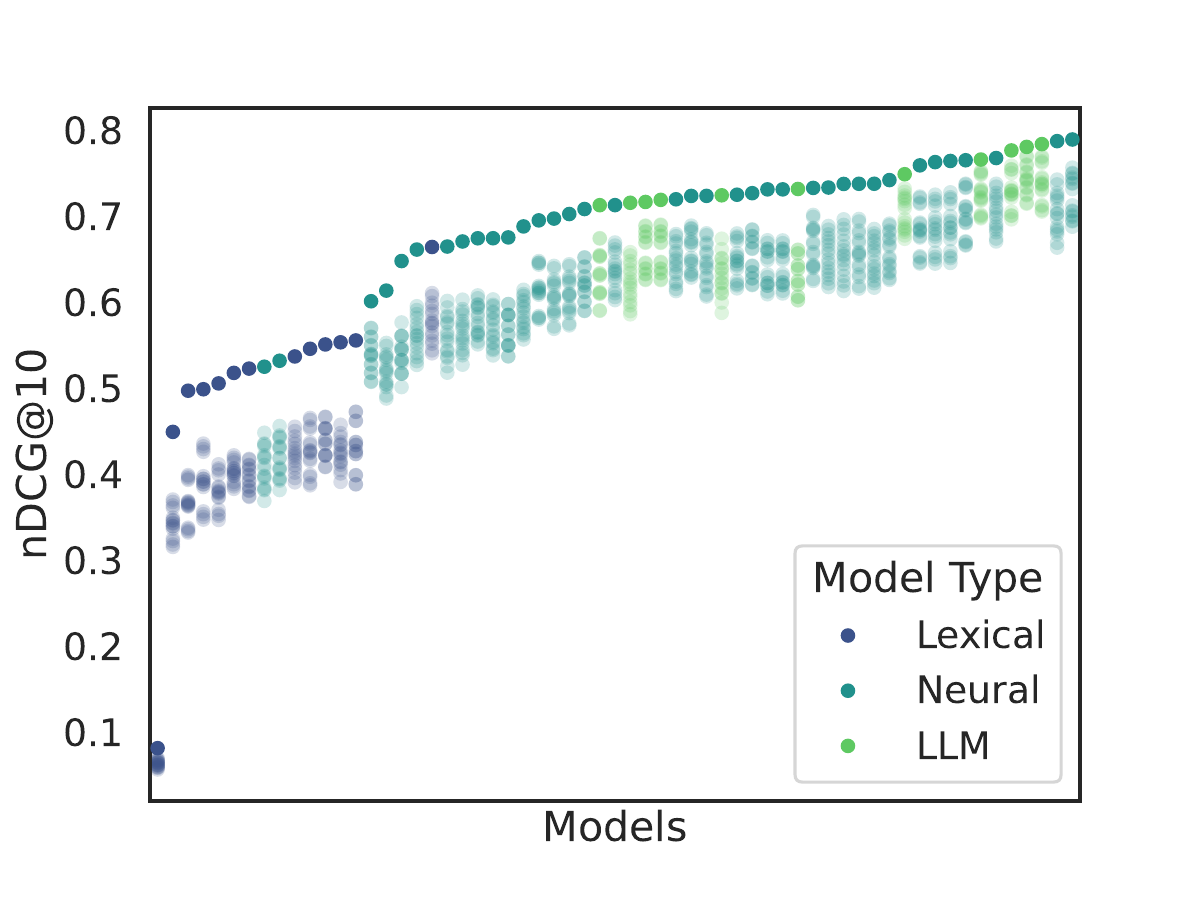}
\caption{Variance in nDCG@10 over annotator combinations. Effectiveness values on new annotations are translucent. Color of points indicates model type.\vspace{.5cm}}
\Description{Shows that generally, nDCG values degrade, suggesting more specific intents expressed in secondary annotations.}
\label{fig:delta}
\end{figure}

\begin{figure*}
    \centering
    \includegraphics[width=1\linewidth]{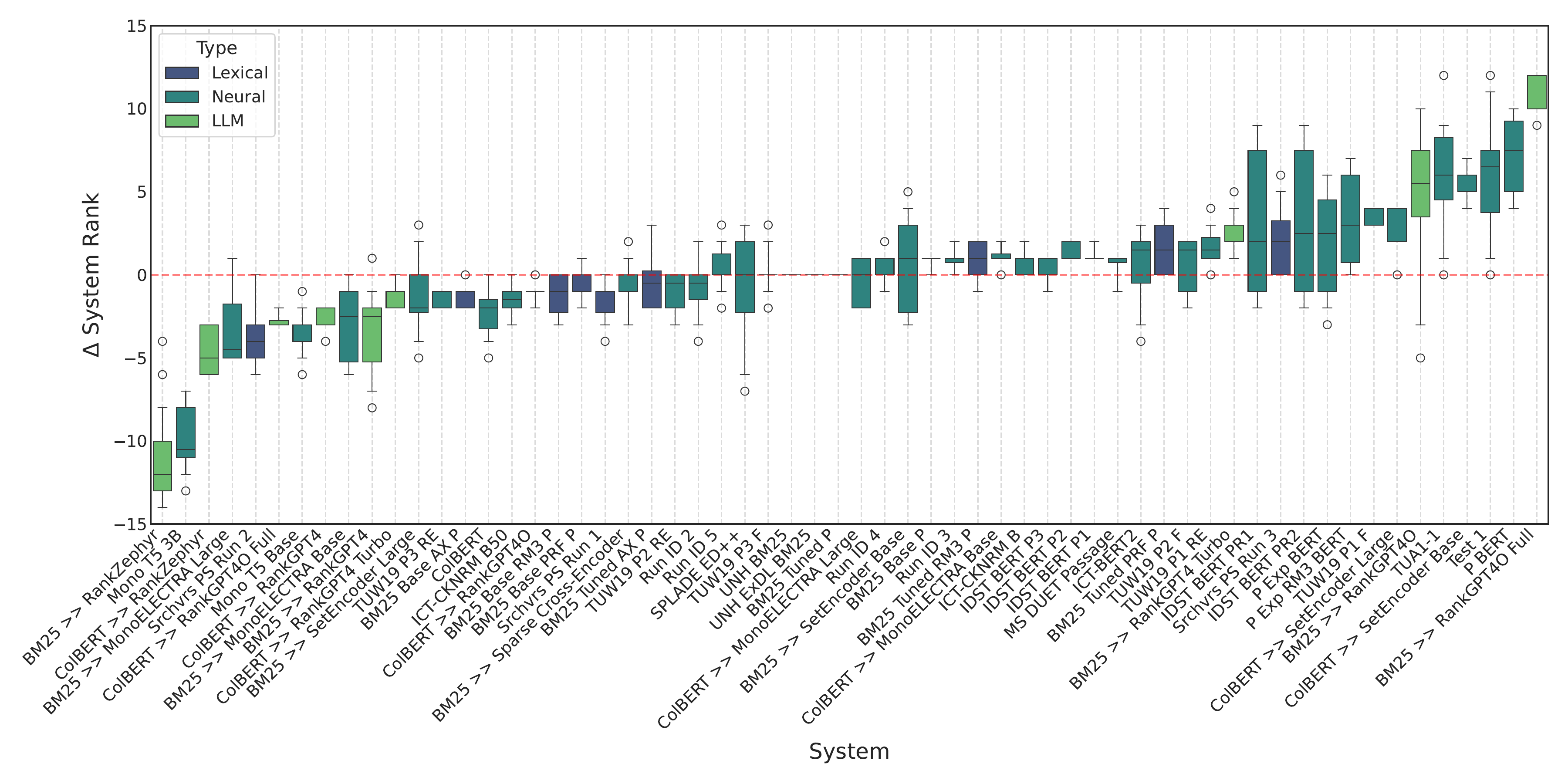}
    \caption{Aggregated system rank changes over all natural combinations of secondary versus primary annotations measured using nDCG@10. Negative $\Delta$ indicates a system frequently improves in rank.}
    \label{fig:delta_rank}
    \Description{Shows that while some more traditional neural systems remain stable, depending on model capacity and distillation source overfitting can occur to the original DL'19 judgments.}
\end{figure*}
\input{src/table/corr}

The core objective in re-annotation studies is to validate that downstream evaluation is not compromised by human bias in subjective relevance judgments. By measuring the stability of system order and the correlation of system order by target metrics, we can assess collection reliability. Following \citet{voorhees:2000} and as described in Section \ref{sec:perm}, we sample combinations of relevance judgments and observe the stability of pair-wise system comparisons in Figure \ref{fig:sample_delta}. Comparisons closest to the top right of the figure represent those where nDCG differences are large; however, they frequently swap in system order. A clear pareto-frontier is composed of comparisons between LLM and neural systems under judgment combinations. Lexical-neural comparisons generally lie in the main body of neural-neural and llm-llm comparisons, suggesting similar stability as was noted by \citet{voorhees:2000} over TREC-6. However, more generally, we replicate the findings of previous works that despite low annotator agreement, under re-annotation, system ordering is highly correlated with the original DL'19 annotations even under 4-grade relevance. In Table \ref{tab:corr}, observe aggregate correlation values using both natural combinations of annotators and the in-sample approach described by \citet{voorhees:2000}.

As the In-Sample approach includes primary judgments, a marginally higher correlation compared to combinations of solely secondary annotators is found. To investigate where system order diverges, we analyse ranking changes at a system level. In Figure \ref{fig:delta} see that generally, new annotations led to a reduction in absolute nDCG@10 values relative to the original annotations, effectiveness measures on original annotations order systems, and that system order is not fully stable but LLM-based or distilled systems appear to be outliers, as such we explicitly measure the changes in system order concentrated around these systems. In Figure~\ref{fig:delta_rank}, we combine all possible combinations of annotators across annotation pairs. We measure changes in system order under these combinations and aggregate by system. Many lexical systems remain stable in ordering ($\Delta\approx0$). Older neural approaches in the original system pool display higher variance but can improve or degrade in rank without a significant difference. When considering only modern systems, we observe that smaller models trained with supervised and semi-supervised approaches consistently improve system rank. Within this group, however, larger semi-supervised models are more robust than their smaller counterparts under new judgments. List-wise encoders generally degrade in rank. These models use an interaction token to facilitate inter-document attention, assisting in out-of-domain effectiveness~\cite{schlatt:2024}. Changes in effectiveness are more pronounced and are significant in smaller distilled language models. This is notable as though we see continuing improvement in smaller models out-of-domain through distillation. One should be cautious that in-domain effectiveness may be less robust than benchmarks indicate, which for many applications may be more important than robust effectiveness across heterogeneous corpora.

\input{src/table/oracles}

\enlargethispage{\baselineskip}
Runs that are either zero-shot or distilled from LLM re-rankers show the highest variance in relative effectiveness even within a sub-group such as those based on GPT 4 endpoints. GPT-4o-based runs significantly degrade in system rank when re-ranking BM25 but are more stable under a neural first-stage ColBERT. This is most likely caused by positional bias found in list-wise ranking towards already precise first-stage rankings~\cite{tang:2024, parry:2024}. This hypothesis is validated by the frequent rank improvement of RankZephyr (the left-most model in Figure \ref{fig:delta}), which was trained explicitly to reduce this bias~\cite{pradeep:2023}. Overall, we consider that though high-quality data (i.e., granular training data from strong models) is important in improving the robustness of neural models, these findings would suggest that smaller models in our current training strategies may not have the capacity to robustly generalize even in-domain.

\subsection{Systems versus Humans}

Recall our argument that a ``perfect'' score on a normalized metric may not be a feasible or desirable goal in ranking evaluation; as such, we measure the effectiveness of human annotators aggregated in multiple forms outlined in Section~\ref{sec:oracle}. In Table~\ref{tab:oracl}, see that systems can exceed or reach parity with secondary annotators, particularly with respect to nDCG@10 and MRR@10, with multiple systems, both LLM and encoder-based, exceeding annotators. When excluding unjudged documents (which is known to overestimate the retrieval effectiveness~\cite{froebe:2023,sakai:2008}), annotator performance measured by precision can be exceeded by neural systems across nDCG, P, and MRR; however, recall remains lower in all settings. Given our findings in Figure~\ref{fig:delta}, an observation that systems are better than humans should be considered cautiously. However, the robustness of the LLM-based RankZephyr observed both by system order robustness over new judgments and its parity with human annotators suggests that with sufficient capacity, a smaller system can be distilled from larger systems while remaining robust. Observing similar effectiveness to humans and confidence intervals providing no way to discriminate systems, we consider that achieving significant improvements would suggest far exceeding humans, which may not be possible in a way conducive to improving information access and may solely serve to improve precision on specific topics and intents. The ability of distilled systems to reach parity with human effectiveness across multiple annotator combinations while obtaining largely reduced recall in a judged setting, even when re-ranking a neural first-stage retriever, is undesirable for several downstream tasks. Although precision is approaching or overtaking the effectiveness of an oracle, recall remains a challenge in densely annotated collections.

%% file: src/table/models.tex
\begin{table}[tb]
    \centering
    \caption{Modern systems included in our analysis. Rightmost columns are the source of labels for semi-supervised learning; neural is any encoder-type model, and LLM is a Large Language Model. Absence of both indicates binary labels.}
    \small
    \setlength\tabcolsep{.5pt}
    \begin{tabular*}{\columnwidth}{@{\extracolsep{\fill}}llc@{\hspace{2\tabcolsep}}c@{}}
    \toprule
         & & \multicolumn{2}{c}{Distillation} \\
    \cmidrule(r@{0pt}){3-4}
        Model & Type & Neural & LLM \\ 
    \midrule
        ColBERT~\cite{khattab:2020} & Late interaction bi-encoder & \texttimes & \texttimes \\
        SPLADE ED++~\cite{formal:2021, formal:2022} & Learned sparse encoder & \checkmark & \texttimes \\
        MonoT5~\cite{nogueira:2020} & Seq2Seq cross-encoder & \texttimes & \texttimes \\ 
        MonoELECTRA~\cite{pradeep:2022, schlatt:2024b} & BERT-based cross-encoder & \checkmark  & \checkmark \\ 
        SetEncoder~\cite{schlatt:2024} & BERT-based list-wise cross-encoder & \checkmark & \checkmark \\
        RankZephyr~\cite{pradeep:2023} & Decoder-only list-Wise cross-encoder & \texttimes & \checkmark \\ 
        RankGPT~\cite{sun:2023} & Decoder-only list-wise cross-encoder & ? & ? \\ 
        \bottomrule
    \end{tabular*}
    \label{tab:models}
\end{table}

%% file: src/table/pilot.tex
\begin{table}[tb]
    \centering
    \setlength{\tabcolsep}{5pt}
    \caption{Comparing the ration of annotation grades between the primary annotators and secondary annotators in a pilot study of 10 topics.}
    \begin{tabular*}{\columnwidth}{@{\extracolsep{\fill}}cccccc@{}}
        \toprule
        & \multicolumn{4}{c}{Grade Ratio} & \\
        \cmidrule{2-5}
        Annotators & $0$ & $1$ & $2$ & $3$ & $|A|$ \\
        \midrule
        Primary & 0.487 & 0.190 & 0.308 & 0.015 & 1143 \\
        Secondary & 0.560  & 0.213 & 0.108 & 0.118 & 600 \\
        \bottomrule
    \end{tabular*}
    \label{tab:pilot}
\end{table}

%% file: src/table/agreement.tex
\begin{table}[tb]
    \centering
    \setlength{\tabcolsep}{5pt}
    \caption{Inter-Annotator agreement over combinations of annotators. Agreement is shown under binarised and 4-grade relevance.}
    \begin{tabular*}{\columnwidth}{@{\extracolsep{\fill}}cccccccc@{}}
        \toprule
        & \multicolumn{2}{c}{Fleiss' $\kappa$} & \multicolumn{2}{c}{Overlap} & \multicolumn{2}{c}{Cohen's $\kappa$} &  \\
        \cmidrule(r@{.8em}){2-3}
        \cmidrule(r@{.8em}){4-5}
        \cmidrule(r@{.8em}){6-7}
        Group & 4 & 2 & 4 & 2 & 4 & 2 & |Q| \\
        \midrule
        Primary+1 & 0.22 & 0.44 & 0.12 & 0.46 & 0.12 & 0.31 & \multirow{2}{*}{14} \\ 
        \cmidrule{1-7}
        1 & --- & --- & 0.42 & 0.72 & 0.19 & 0.37 &  \\ 
        \midrule
        Primary+2 & 0.28 & 0.40 & 0.17 & 0.48 & 0.17 & 0.31 & \multirow{2}{*}{12} \\ 
        \cmidrule{1-7}
        2 & --- & --- & 0.47 & 0.74 & 0.22 & 0.38 &  \\ 
        \midrule
        Primary+3 & 0.17 & 0.31 & 0.11 & 0.45 & 0.11 & 0.28 & \multirow{2}{*}{8} \\ 
        \cmidrule{1-7}
        3 & --- & --- & 0.63 & 0.89 & 0.27 & 0.47 &  \\ 
        \midrule
        Primary+4 & 0.28 & 0.37 & 0.19 & 0.48 & 0.19 & 0.30 & \multirow{2}{*}{9} \\ 
        \cmidrule{1-7}
        4 & --- & --- & 0.43 & 0.71 & 0.19 & 0.34 &  \\ 
        \bottomrule
    \end{tabular*}
    \label{tab:agree}
\end{table}

%% file: src/table/corr.tex
\begin{table}[tb]
    \centering
    \setlength{\tabcolsep}{4pt}
    \caption{System ranking correlation over different annotator groups measuring Kendall's $\tau$, Spearmans's $\rho$ and rank-biased overlap. `Combination' is the mean of natural permutations of the 8 annotators, `In-Sample' is the mean of judgment permutations as described by \citet{voorhees:2000}. All $\tau$ and $\rho$ values represent significant correlation.}
    \begin{tabular*}{\columnwidth}{@{\extracolsep{\fill}}p{4cm}lll@{}}
        \toprule
        Type & $\tau$ & $\rho$ & RBO \\
        \midrule
        Combination & 0.879 & 0.972 & 0.888 \\
        In-Sample & 0.897 & 0.977 & 0.902 \\
        \bottomrule
    \end{tabular*}
    \label{tab:corr}
\end{table}

%% file: src/table/oracles.tex
\newcommand{\allphantom}{\phantom{^{\textbf{ABC}}}}
\newcommand{\phantomone}{\phantom{A}}
\newcommand{\phantomtwo}{\phantom{BC}}

\begin{table}[tb]
    \centering
    \caption{Comparing aggregate annotations as rankings with several ranking architectures. Judged Only indicates that original runs have been filtered to contain solely annotated documents, allowing for a similar setting to the annotator rankings. Significance is with respect to Minimum (\textbf{A}), Mean (\textbf{B}) or Maximum (\textbf{C}) aggregation via a paired t-test with Bonferroni correction ($p<0.05$).}
    \footnotesize
    \setlength\tabcolsep{0pt}
    \begin{tabular*}{\columnwidth}{@{\extracolsep{\fill}}lp{1.5cm}p{1.5cm}p{1.5cm}p{1.5cm}}
        \toprule
        Model & nDCG@10 & P@10 & MRR@10 & R@100 \\
        \midrule
        \multicolumn{5}{l}{Judged \& Unjudged} \\
        \cmidrule{1-1}
        BM25 & $0.51 \pm 0.08^{\textbf{ABC}}$ & $0.41 \pm 0.09^{\textbf{ABC}}$ & $0.70 \pm 0.12^{\textbf{BC\phantomone}}$ & $0.49 \pm 0.10^{\textbf{ABC}}$ \\ 
        Splade ED++ & $0.73 \pm 0.07^{\textbf{BC\phantomone}}$ & $0.62 \pm 0.10^{\textbf{BC\phantomone}}$ & $0.91 \pm 0.07\allphantom$ & $0.60 \pm 0.09^{\textbf{ABC}}$ \\
        ColBERT>>RankZephyr & $0.75 \pm 0.07^{\textbf{B\phantomtwo}}$ & $0.67 \pm 0.10\allphantom$ & $0.84 \pm 0.09\allphantom$ & $0.67 \pm 0.09^{\textbf{ABC}}$ \\ 
        ColBERT>>RankGPT-4o & $0.78 \pm 0.06\allphantom$ & $0.71 \pm 0.09\allphantom$ & $0.87 \pm 0.08\allphantom$ & $0.67 \pm 0.09^{\textbf{ABC}}$ \\ 
        \midrule
        \multicolumn{5}{l}{Judged Only} \\
        \cmidrule{1-1}
        BM25 & $0.51 \pm 0.08^{\textbf{ABC}}$ & $0.41 \pm 0.09^{\textbf{ABC}}$ & $0.70 \pm 0.12^{\textbf{BC\phantomone}}$ & $0.65 \pm 0.09^{\textbf{ABC}}$ \\ 
        Splade ED++ & $0.73 \pm 0.07^{\textbf{B\phantomtwo}}$ & $0.63 \pm 0.10^{\textbf{BC\phantomone}}$ & $0.91 \pm 0.07\allphantom$ & $0.69 \pm 0.09^{\textbf{BC\phantomone}}$ \\
        ColBERT>>RankZephyr & $0.77 \pm 0.07\allphantom$ & $0.70 \pm 0.09\allphantom$ & $0.85 \pm 0.08\allphantom$ & $0.67 \pm 0.09^{\textbf{ABC}}$ \\ 
        ColBERT>>RankGPT-4o & $0.80 \pm 0.05\allphantom$ & $0.73 \pm 0.09\allphantom$ & $0.89 \pm 0.08\allphantom$ & $0.67 \pm 0.09^{\textbf{ABC}}$ \\ 
        \midrule
        Minimum (\textbf{A})& $0.76 \pm 0.07\allphantom$ & $0.67 \pm 0.10\allphantom$ & $0.84 \pm 0.09\allphantom$ & $0.75 \pm 0.06\allphantom$ \\ 
        Mean (\textbf{B})& $0.81 \pm 0.05\allphantom$ & $0.71 \pm 0.10\allphantom$ & $0.90 \pm 0.08\allphantom$ & $0.86 \pm 0.06\allphantom$ \\
        Maximum (\textbf{C})& $0.79 \pm 0.05\allphantom$ & $0.70 \pm 0.09\allphantom$ & $0.86 \pm 0.08\allphantom$ & $0.86 \pm 0.06\allphantom$ \\ 
        \bottomrule
    \end{tabular*}
    \label{tab:oracl}
\end{table}

%% file: sigir-25-dl-annotations-sum.tex
\section{Conclusion}

We considered important studies on the impact of the annotator agreement on retrieval evaluations in light of modern neural evaluation scenarios. In reproducing re-annotation processes and evaluations by \citet{voorhees:2000} on a modern test collection, we validate the stability of system ordering under re-annotation under 4-grade relevance and ambiguous query intent. We raised concerns that the broad and frequent use of test collections to ``tune'' learned systems implicitly through influences from published works may reduce the reliability of a collection over time. We identified that the subjectivity of relevance through query intent may be a factor in determining ``overfitting'' in ranking tasks. As a query may have several interpretations, we re-annotate the popular TREC DL'19 collection. We found that system order varies when evaluating large language models and systems distilled from them. Furthermore, we observed that the current state-of-the-art can outperform combinations of human annotators on original relevance judgments, suggesting we may have reached a realistic bound on precision for this collection. We posit that further improvement on this collection may not indicate that a system is better than another in a meaningful way. Our process has limitations---most notably, re-annotation is incredibly costly. Future work will investigate how to draw similar conclusions about a collection's expiration without the need for extensive human effort.

\begin{acks}
We thank all involved in making the sessions of the Collab-a-thon at ECIR`24~\cite{macavaney:2024} an engaging space for collaboration.
\end{acks}